# Electromotive field in space and astrophysical plasmas

Ph.-A. Bourdin[1] · Y. Narita[2]

**Abstract**
The concept of electromotive field appears in various applications in space and astrophysical plasmas. A review is given on the electromotive field highlighting our current understanding of the theoretical picture and the spacecraft observations in interplanetary space. The electromotive field is a key concept to successfully close the set of turbulent magnetohydrodynamic equations and also to construct a more complete picture of space plasma turbulence. Applications to astrophysical cases (Earth magnetosphere, heliospheric shocks, interstellar medium, and relativistic jets) are also briefly introduced, as well.

## 1 Introduction

Various space and astrophysical bodies are known to exhibit large-scale magnetic fields from the planetary scale to the stellar and interstellar scale, and further to the galactic scale. The dynamo mechanism is considered as the likely candidate to explain the origin of the large-scale magnetic field, generating coherent structures out of small-scale turbulent motions. The generated large-scale field in turn develops into turbulence by nonlinearities in the fluid and plasma.

The question is addressed here, "What determines the evolution of space and astrophysical plasmas into construction of the large-scale magnetic field by the dynamo mechanism or into destruction of the large-scale field into turbulence?" The concept of the electromotive field (abbreviated hereafter as EMF, often referred to as the electromotive field in the literatures) is the key concept to answer the question. The EMF formulates the magnetic field evolution as the competition between the construction effect by the dynamo mechanism and the destruction effect by plasma turbulence. The EMF is regarded as one of the electric field realizations in electrically conducting fluids or plasmas. Since the EMF is defined as the cross product between the fluctuating flow velocity and the fluctuating magnetic field, one can evaluate the EMF without directly measuring the electric field.

✉ Ph.-A. Bourdin
  Philippe.Bourdin@uni-graz.at

✉ Y. Narita
  y.narita@tu-braunschweig.de

[1] Institut für Physik, Karl-Franzens-Universität, Universitätsplatz 5, 8010 Graz, Austria
[2] Institut für Theoretische Physik, Technische Universität Braunschweig, Mendelssohnstr. 3, 38106 Braunschweig, Germany

The EMF is the key concept in the mean-field dynamo model in which the large-scale magnetic field is generated by amplifying small-scale magnetic fields in turbulent fluid motions. The original idea of the dynamo mechanism was developed by Elsasser in the 1950s (Elsasser 1956) and Steenbeck, Krause, and Rädler in the 1960s (Steenbeck et al. 1966). Later on, the concept was further elaborated and disseminated by Roberts and Stix (1971), Moffatt (1974, 1978), Parker (1979), Krause and Rädler (1980), and Roberts and Soward (1992).

Examples of the large-scale magnetic field presumably caused by the dynamo mechanism can be found in the solar system such as the Earth (Glatzmaier and Roberts 1998; Glatzmaier 2002; Roberts and Glatzmaier 2000; Kono and Roberts 2002), the planets (except for Venus and Mars) (Jones 2011), Ganymede (Jupiter satellite) (Schubert et al. 1996; Sarson et al. 1997), and the Sun (Charbonneau 2010, 2014; Brandenburg 2018). Extrasolar magnetic fields can be found at the stars (e.g., pre main sequence starts, low-mass stars, solar-like stars) (Berdyugina 2005; Brun and Browning 2017), and in the galactic and extragalactic space (Vainshtein and Ruzmaikin 1971; Kronberg 1994; Widrow 2002; Beck et al. 2020).

Our understanding of the dynamo mechanism is being deepened and broadened with the help of numerical simulations solving the fundamental equations directly as well as analytic treatment and modeling (Brandenburg 2018). Recent theoretical studies by Yokoi (2018a) and Yokoi and Tobias (2021) suggest that the EMF can substantially be influenced by the density variation, and the EMF is locally enhanced such as in the shock-front region. The density enhancement leads to even a fast magnetic reconnection. Motivated by the pioneering works by Marsch and Tu (1992, 1993), recent precursor studies using the Helios data in the solar wind show that the EMF is excited during the passage of magnetic cloud or interplanetary shocks (Bourdin et al. 2018; Narita and Vörös 2018; Hofer and Bourdin 2019).

This article is an update of the recent review by Narita (2021), extending the review on the following points: (1) theoretical picture of the EMF, (2) new insights from reinterpretation of the earlier observations and new observations, and (3) EMF in astrophysical systems. The concept of EMF can be implemented in the spacecraft data to construct a more complete picture of the turbulent fluctuations in space. It is believed that the research field of EMF is one of the likely candidates of breakthrough in space and astrophysical plasma physics.

## 2 Theoretical picture

### 2.1 A simplified view

The EMF is constructed as the statistically averaged vector product between the fluctuating flow velocity $\delta U$ and the fluctuating magnetic field $\delta B$ as

$$\mathcal{E} = \langle \delta U \times \delta B \rangle. \tag{1}$$

In our notation, the flow velocity $U$ and the magnetic field $B$ are decomposed into the mean-field (or large-scale field) denoted by a subscript null and a fluctuating-field part (or small-scale field) denoted by delta in front of the field

$$U = U_0 + \delta U \tag{2}$$

$$B = B_0 + \delta B. \tag{3}$$

The large-scale field represents a smooth field, and its fluctuation is assumed to vanish. Also, the large-scale (or mean value) of the fluctuation field vanishes in this picture. It is obvious in Eq. (1) that a linearly polarized quasi-monochromatic Alfvén wave in the MHD picture cannot excite the EMF, as the quasi-monochromatic Alfvén wave has either a phase-synchronized oscillation between the fluid motion and the wave magnetic field (when propagating anti-parallel to the mean magnetic field) or a phase-anti-synchronized oscillation when propagating parallel to the mean field. The EMF has the units of electric field such as V m$^{-1}$. For example, typical values in the solar wind are a velocity fluctuation of about 1 km s$^{-1}$ and a magnetic field fluctuation of about 1 nT. The EMF is of the order of 1 mV km$^{-1}$.

The EMF in the form of Eq. (1) is smoothly derived by applying the field decomposition (Eqs. 2 and 3) to the induction equation

$$\partial_t B_0 = \nabla \times (U_0 \times B_0) + \nabla \times \langle \delta U \times \delta B \rangle + \eta \nabla^2 B_0. \tag{4}$$

Here, $\eta$ is the magnetic diffusivity, which is related to the conductivity $\sigma$ through $(\mu_0 \sigma)^{-1}$. The first term on the right hand side in Eq. (4) represents the frozen-in large-scale magnetic field, the second term the curl of EMF, and the third term the diffusion of large-scale field. The EMF can act both as constructive to the large-scale field (e.g., amplification of large-scale field by fluctuations such as in the dynamo mechanism) and as destructive (e.g., scattering or disturbance of large-scale field by fluctuations such as in plasma turbulence)

In the simplified mean-field electrodynamics, it is assumed that the EMF depends on the large-scale magnetic field and it spatial derivative (the curl operator) in the form of

$$\mathcal{E} = \alpha B_0 + \beta \nabla \times B_0, \tag{5}$$

where the transport coefficient $\alpha$ denotes the growth or damping (depending on the sign) of the large-scale magnetic field and $\beta$ the turbulent diffusivity. It is illustrative that if the both transport coefficients $\alpha$ and $\beta$ are regarded as constant, the induction equation (Eq. 4) becomes

$$\partial_t \boldsymbol{B}_0 = \nabla \times (\boldsymbol{U}_0 \times \boldsymbol{B}_0) + \alpha \nabla \times \boldsymbol{B}_0 + (\beta + \eta)\nabla^2 \boldsymbol{B}_0. \tag{6}$$

That is, the alpha term in the EMF (Eq. 5) gives the possibility of an exponential growth of the large-scale magnetic field (if $\alpha$ is a positive constant), and the beta term enhances the diffusion of the large-scale magnetic field by turbulent diffusion (or scattering). It is useful to note that the Ansatz form of EMF (Eq. 5) can analytically be derived when treating interacting and colliding Alfvén waves (Namikawa and Hamabata. 1982b).

Evaluation of the transport coefficients needs the knowledge on the individual turbulence realizations, that is, the statistical properties and geometric configuration of the small-scale turbulent fields determine if the large-scale field grows or damps. Historically speaking, the following form of the transport coefficients was introduced in the 1960s (Steenbeck et al. 1966; Krause and Rädler 1980)

$$\alpha = -\frac{1}{3}\tau \langle \delta \boldsymbol{U} \cdot (\nabla \times \delta \boldsymbol{U}) \rangle \tag{7}$$

$$\beta = \frac{1}{3}\tau \langle \delta \boldsymbol{U} \cdot \delta \boldsymbol{U} \rangle. \tag{8}$$

Here, a proper time scale $\tau$ (turbulence correlation time) needs to be assessed in addition such as the eddy turnover time. Equation (7) indicates that the helical sense (or the field-rotation sense) of the fluid motion drives the dynamo action, and Eq. (8) indicates that the turbulent fluctuation is a source of effective diffusion (turbulent diffusion) to the large-scale magnetic field. (Namikawa and Hamabata. 1982b) analytically obtained the transport coefficients for interacting Alfvén waves, and the alpha coefficient is proportional to the residual helicity (difference between the kinetic helicity and the current helicity).

The EMF completes the set of second-order fluctuation quantities such as energy and helicity densities (hereafter, we omit "densities" for brevity). The EMF is directly accessible by computing the cross product of the fluctuating flow velocity and the fluctuating magnetic field. From the experimental point of view, one may construct the covariance matrix $\mathbf{R}_{ub}$ (or cross spectral density matrix in the Fourier domain) using the flow velocity and the magnetic field, and the EMF appears as off-diagonal elements of the following matrix:

$$\mathbf{R}_{ub} = \begin{bmatrix} \langle \delta U_x \delta B_x \rangle & \langle \delta U_x \delta B_y \rangle & \langle \delta U_x \delta B_z \rangle \\ \langle \delta U_y \delta B_x \rangle & \langle \delta U_y \delta B_y \rangle & \langle \delta U_y \delta B_z \rangle \\ \langle \delta U_z \delta B_x \rangle & \langle \delta U_z \delta B_y \rangle & \langle \delta U_z \delta B_z \rangle \end{bmatrix}.. \tag{9}$$

The x component of the EMF is, for example, evaluated as the difference between the y–z element and its transposed element of the matrix

$$\mathcal{E}_x = \langle \delta U_y \delta B_z \rangle - \langle \delta B_y \delta U_z \rangle. \tag{10}$$

It is worth mentioning that the trace of the covariance matrix, tr($\mathbf{R}_{ub}$) represents the (averaged) cross helicity $h_{crs}$ of the fluctuations

$$h_{crs} = \langle \delta U_x \delta B_x \rangle + \langle \delta U_y \delta B_y \rangle + \langle \delta U_z \delta B_z \rangle, \tag{11}$$

which is an invariant of ideal magnetohydrodynamics when integrated over a proper volume.

The transport coefficients such as $\alpha$ and $\beta$ are expected to depend on the second-order fluctuation quantities such as the energy and helicity. For example, the $\alpha$ coefficient in Eq. (7) is proportional to the kinetic helicity

$$h_{kin} = \langle \delta \boldsymbol{U} \cdot (\nabla \times \delta \boldsymbol{U}) \rangle, \tag{12}$$

and the $\beta$ coefficient in Eq. (8) is proportional to the flow kinetic energy

$$e_{kin} = \langle \delta \boldsymbol{U} \cdot \delta \boldsymbol{U} \rangle. \tag{13}$$

The build-up of large-scale magnetic field in a helical flow is demonstrated using a semi-analytic and numerical study of magnetohydrodynamic turbulence. In the context of the turbulence inverse cascade (structure formation), Pouquet et al. (1976) showed that the difference between the kinetic helicity and the current helicity (referred to as the residual helicity) essentially contributes to the dynamo mechanism.

In the Fourier domain, since the nabla operator is simply replaced by the wavevector as $\nabla \to i\boldsymbol{k}$, the kinetic helicity is evaluated from the off-diagonal elements of the spectral matrix for the flow velocity $\mathbf{R}_{uu}$. The full spectral matrix $\mathbf{R}_{uu}$ determine the off-diagonal elements of the matrix $\mathbf{R}_{ub}$ in the simplified model. The cross helicity $h_{crs}$ (trace of $\mathbf{R}_{ub}$), the current helicity $h_{crt} = \langle \delta \boldsymbol{B} \cdot (\nabla \times \delta \boldsymbol{B}) \rangle$ (off-diagonal elements of the magnetic field spectral matrix $\mathbf{R}_{bb}$ in the Fourier domain), the magnetic helicity $h_{mag} = \langle \delta \boldsymbol{A} \cdot \delta \boldsymbol{B} \rangle$ (also off-diagonal elements of $\mathbf{R}_{bb}$ in the Fourier domain when the vector potential is obtained from the magnetic field by the uncurling procedure), and the magnetic energy $e_{kin} = \langle \delta \boldsymbol{B} \cdot \delta \boldsymbol{B} \rangle$ (trace of the matrix $\mathbf{R}_{bb}$) are irrelevant in the simplified view of dynamo mechanism described by Eqs. (5)–(8). Magnetic helicity, for example, describes the three-dimensional topological properties of magnetic field lines (Berger and Field 1984; Berger 1999). Helical structures also play an important role in fluid dynamics (Moffatt 2014). Efforts have been put to construct a more comprehensive picture of the EMF (i.e., dependence of the electromotive field on the large-scale fields and turbulence properties) by extending the simplified model in various ways.

### 2.2 Alfvén wave model

Analytic properties of the EMF have extensively studied in the 1980s for counter-propagating Alfvén waves in the incompressible collisionless plasma with zero cross helicity (Namikawa and Hamabata 1982a; Namikawa and Hamabata. 1982b). The alpha term in the EMF is derived as

$$\mathcal{E} = \alpha \boldsymbol{B}_0, \tag{14}$$

and, moreover, the alpha coefficient is determined by the difference between the kinetic helicity and the current helicity in the integration over the wavenumbers

$$\alpha = -\int_0^\infty dk \, (\tilde{h}_{\text{kin}} - \tilde{h}_{\text{crt}}) \frac{\sin(2V_A tk)}{2V_A k}. \tag{15}$$

Here, the helicities with the tilde symbol are the Fourier-transformed quantities defined as for example

$$h_{\text{kin}} = \int_0^\infty dk \, \tilde{h}_{\text{kin}}. \tag{16}$$

In a current-free helical flow with the following helicity spectrum:

$$\tilde{h}_{\text{kin}} = \tilde{h}_{\text{kin},0} \frac{\exp(-2k/k_0)}{(k/k_0)^2} \tag{17}$$

$$\tilde{h}_{\text{crt}} = 0, \tag{18}$$

the alpha coefficient can analytically be evaluated as

$$\alpha = -\frac{\tilde{h}_{\text{kin},0}}{4V_A} \frac{V_A k_0 t}{[1 + (V_A k_0 t)^2]^2}. \tag{19}$$

If the kinetic helicity peaks on a larger scale such that a condition of $V_A k_0 t < 1$ (i.e., $k_0$ is sufficiently small) holds, the expression of the alpha coefficient restores that of the simplified view except for the difference in the factor 1/3

$$\alpha \simeq -\frac{1}{4} h_{\text{kin}} t, \tag{20}$$

where the kinetic helicity is approximated as $h_{\text{kin}} \simeq \tilde{h}_{\text{kin}} k_0$.

The beta effect is introduced by expanding the mean magnetic field to the first order of spatial derivative as (Moffatt 1978; Namikawa and Hamabata. 1982b)

$$B_{0,i} = \sum_k^{x,y,z} x_k \partial_k B_{0,i} \tag{21}$$

$$= x \partial_x B_{0,i} + y \partial_y B_{0,i} + z \partial_z B_{0,i}. \tag{22}$$

The EMF by collecting the alpha and beta effects is

$$\mathcal{E} = \alpha \boldsymbol{B}_0 - \beta_0 \nabla \times \boldsymbol{B}_0 - \beta_1 \delta \boldsymbol{B} \times (\delta \boldsymbol{B} \cdot \nabla) \boldsymbol{B}_0. \tag{23}$$

The field-aligned EMF excited by the Alfvén waves can be of the order of several kV in space, and is a candidate mechanism of auroral electron acceleration during the geomagnetic substorms (Namikawa and Hamabata 1982a; Namikawa and Hamabata. 1982b; Namikawa et al. 1982). The interacting Alfvén waves may form standing Alfvén waves such as in the Earth magnetic field, and it was shown that the alpha effect can be generated by the standing Alfvén waves if the initial wave fields have non-zero helicity (Namikawa and Hamabata 1988).

The EMF in a collisionless non-uniform plasma is presumed to depend not only on the mean magnetic field and its spatial gradient, but also on the velocity shear and the density gradient as (Namikawa and Hamabata 1982a; Hamabata et al. 1982)

$$\begin{aligned}\mathcal{E} =& \alpha_1 \nabla_\parallel \boldsymbol{B}_0 + \alpha_2 \nabla_\perp \boldsymbol{B}_0 + \alpha_3 \boldsymbol{B}_0 \times (\nabla \times \boldsymbol{B}_0) + \\ & \beta_1 \nabla_\parallel U_{0,\parallel} + \beta_2 \nabla_\perp U_{0,\parallel} + \beta_3 \boldsymbol{B}_0 \times (\nabla \times \boldsymbol{B}_0) + \\ & \gamma_1 \nabla_\parallel \rho_0 + \gamma_2 \boldsymbol{B}_0 (\boldsymbol{U}_0 \cdot \nabla) \rho + \gamma_3 \nabla_\perp \rho_0.\end{aligned} \quad (24)$$

On the other hand, no impact of the Hall current is expected on the EMF in the Alfvén wave model (Hamabata et al. 1982).

It is also worth mentioning that the ponderomotive force arises from the wave pressure through the Reynolds and Maxwell stress

$$\boldsymbol{F}_\mathrm{pon} = \nabla \cdot \left\langle \frac{1}{\mu_0 \rho} \delta \boldsymbol{B} \delta \boldsymbol{B} - \delta \boldsymbol{U} \delta \boldsymbol{U} \right\rangle. \quad (25)$$

Like the EMF, in the case of zero cross helicity, the ponderomotive force generated by the Alfvén waves (propagating along the mean magnetic field) is fully determined by the energy and helicity spectra. Moreover, it is possible to express the ponderomotive force as a function of the mean magnetic field and mean velocity. The ponderomotive force acts in the perpendicular direction to the mean field (Namikawa and Hamabata 1983). For example, the simplest form of ponderomotive force generated by parallel-propagating waves is

$$\boldsymbol{F}_\mathrm{pon} = -\frac{1}{2} \beta^\pm \{ \pm B_0/V_\mathrm{A} \hat{\boldsymbol{b}} \times \nabla[\hat{\boldsymbol{b}} \cdot \nabla(\hat{\boldsymbol{b}} \cdot \boldsymbol{B}_0)] + \hat{\boldsymbol{b}} \times \nabla[\hat{\boldsymbol{b}} \cdot \nabla(\hat{\boldsymbol{b}} \cdot \boldsymbol{U}_0)] \}, \quad (26)$$

where

$$\beta^\pm = \pm V_\mathrm{A}^{-1} \int_0^\infty \mathrm{d}k \, \frac{\tilde{h}_\mathrm{kin}^\pm(k)}{k^2} \quad (27)$$

are the coefficients associated with the helical flow sense for the parallel-propagating waves for the plus sign (at positive frequencies) and the anti-parallel-propagating waves for the minus sign (at negative frequencies). More extended expressions of the Ponderomotive force are presented in Namikawa and Hamabata (1983) and Hamabata and Namikawa (1988). The contribution of the current helicity to the alpha effect was also analytically derived by Seehafer (1994).

The transport coefficients $\alpha$ and $\beta$ are evaluated in a variety of turbulence setups by Rädler and Rheinhardt (2007). For homogeneous isotropic turbulence, the coefficients are evaluated explicitly using the Green function, and are obtained in a simple way as

$$\alpha = -\frac{\tau}{3}\left[\langle \delta \boldsymbol{U} \cdot (\nabla \times \delta \boldsymbol{U})\rangle - \frac{1}{\rho_0 \mu_0}\langle \delta \boldsymbol{B} \cdot (\nabla \times \delta \boldsymbol{B})\rangle\right] \quad (28)$$

$$\beta = \frac{1}{3}\langle \delta \boldsymbol{U}^2 \rangle \tau. \quad (29)$$

### 2.3 Cross-helicity dynamo model

While the picture of the alpha effect is becoming well established and regularly observed in simulations, the dynamo mechanism beyond the alpha effect may possibly exist. For instance, a three-dimensional helical flow motion is not expected in the inner region of accretion disks. Numerical simulations still demonstrate the existence of dynamo action despite the fact that the alpha effect is forbidden in the accretion disks for the symmetry argument reason (Hawley et al. 1995; Lesur and Ogilvie 2008). Moreover, nonlinear effects quench the alpha dynamo before the large-scale magnetic field reaches a significant amplitude (Kulsrud and Anderson 1992; Gruzinov and Diamond 1994). The gradients of flow velocity and higher-order structure of the large-scale magnetic field are likely candidates for a more complete picture of dynamo action.

The EMF may be extended to include the large-scale flow effect through first-order derivative. By doing so, the symmetry is restored between the magnetic field and the flow velocity in the EMF formulation. The concept of large-scale flow effect on the EMF was originally proposed in study of reversed field pinch (Yoshizawa 1990). The concept is further being extended to establishing the cross-helicity dynamo (Yokoi and Balarac 2011; Yokoi 2013). The EMF in the cross-helicity dynamo is formulated as

$$\boldsymbol{\mathcal{E}} = \alpha \boldsymbol{B}_0 - \beta \nabla \times \boldsymbol{B}_0 + \gamma \nabla \times \boldsymbol{U}_0. \quad (30)$$

The first term with the coefficient $\alpha$ represents the amplification of the large-scale magnetic field by the combination of the helical flow motion with the helicity current structure (cf. alpha dynamo mechanism). The second term with the coefficient $\beta$ represents the scattering of the large-scale field by turbulent fluctuations. Again, the $\beta$ term yields $\beta \nabla^2 \boldsymbol{B}_0$ in the induction equation for the large-scale field, which is interpreted as the turbulent diffusion. The third term with the coefficient $\gamma$ represents the amplification of the large-scale magnetic field by the non-zero cross-helicity effect. Note that the coefficient gamma in Eq. (30) is different from that used in the Alfvén wave model. With the cross-helicity model (Eq. 30), the induction equation for the large-scale field reads

$$\partial_t \boldsymbol{B}_0 = \nabla \times (\boldsymbol{U}_0 \times \boldsymbol{B}_0) + \nabla \times (\alpha \boldsymbol{B}_0 + \gamma \nabla \times \boldsymbol{U}_0) \\ + (\beta + \eta)\nabla^2 \boldsymbol{B}_0. \quad (31)$$

The coefficient $\gamma$ is modeled in the same fashion as the coefficients $\alpha$ and $\beta$

$$\gamma = \frac{1}{3}\tau \langle \delta U \cdot \delta B \rangle \tag{32}$$

when simplified, e.g., in Bourdin et al. (2018). The coefficient values are expected to be of the order of $C_\alpha = O(10^{-2})$, $C_\beta = O(10^{-1})$ and $C_\gamma = O(10^{-1})$ (Yokoi 2013).

The cross-helicity dynamo model is successful in reproducing the time-series profile (or the wave form) of the EMF measured by the Helios spacecraft in the solar wind (Bourdin et al. 2018). In contrast, the test for the alpha effect in the EMF (i.e., test for the proportionality between the EMF and the large-scale magnetic field) without the beta or gamma term fails against the solar wind data (Marsch and Tu 1992). The scaling analysis using the Helios data indicates that the alpha, beta, and gamma effects all contribute to the EMF at the same order (Bourdin et al. 2018). Under which condition the alpha or gamma effect will dominate remains an open question.

### 2.4 Effects of higher orders

#### 2.4.1 Rotating body and magnetic shear

Consideration of rotating body and magnetic shear introduces various kinds of new terms in the EMF (Urpin 2002; Rädler et al. 2003; Rädler and Stepanov 2006; Squire and Bhattacharjee 2015). The zeroth-order terms are proportional to the large-scale magnetic field, and the extended version includes the vectorial coupling of the magnetic field with the rotation of the body and the vortical motion of the flow such as $\mathbf{\Omega} \times \boldsymbol{B}_0$ (rotational coupling) $\boldsymbol{\omega} \times \boldsymbol{B}_0$ (vortical coupling) and the component-wise coupling among the magnetic field, the rotation, and the vorticity (6 possible combinations). Here $\mathbf{\Omega}$ is the angular velocity of a rotating body, and $\boldsymbol{\omega} = \nabla \times \boldsymbol{U}$ is the vorticity of the large-scale flow. The first-order terms include the spatial derivative of the large-scale field, such as the current density $\boldsymbol{j}$, the shear of the large-scale field $\nabla \boldsymbol{B}$, and the coupling with the rotation $\mathbf{\Omega}$, and the vorticity $\boldsymbol{\omega}$

$$\mathcal{E}_{\text{alpha}} = \alpha_1 \boldsymbol{B}_0 + \alpha_2 \mathbf{D} \cdot \boldsymbol{B}_0 + \alpha_3 \mathbf{\Omega} \times \boldsymbol{B}_0 + \alpha_4 \boldsymbol{\omega} \times \boldsymbol{B}_0. \tag{33}$$

The magnetic shear (first-order derivative of large-scale field) enters the beta effect in various ways (Rädler and Stepanov 2006; Squire and Bhattacharjee 2015)

$$\begin{aligned}\mathcal{E}_{\text{beta}} = &-\beta_1 \nabla \times \boldsymbol{B}_0 - \beta_2 \mathbf{D} \cdot (\nabla \times \boldsymbol{B}_0) - \beta_3 \mathbf{\Omega}_{\text{rot}} \times (\nabla \times \boldsymbol{B}_0) \\ &- \beta_4 \mathbf{\Omega}_{\text{vor}} \times (\nabla \times \boldsymbol{B}_0) - \beta_5 \mathbf{\Omega}_{\text{rot}} \cdot (\nabla \boldsymbol{B}_0)^{(s)} \\ &- \beta_6 \mathbf{\Omega} \cdot (\nabla \boldsymbol{B}_0)^{(s)} - \beta_7 \epsilon \mathbf{D} \cdot (\nabla \boldsymbol{B}_0)^{(s)},\end{aligned} \tag{34}$$

where $\boldsymbol{D}$ and $(\nabla \boldsymbol{B}_0)^{(s)}$ are the symmetric part of the velocity shear tensor $\nabla \boldsymbol{U}_0$ and the magnetic field shear $\nabla \boldsymbol{B}_0$, respectively. The coupling of the vorticity with the large-scale magnetic shear is analytically derived by Urpin (1999, 2002) for a general turbulent flow.

By considering a mean vortical motion, Urpin (1999) and Rogachevskii and Kleeorin (2003) proposed a contribution of the vorticity to the EMF. The vorticity effect is characterized by the term $\boldsymbol{\omega} \times \boldsymbol{J}_0$), and this was successfully confirmed in the framework of the second-order correlation approximation by Rüdiger and Küker (2016).

### 2.4.2 Hall effect

The Hall effect appears when the ions and the electrons move differently in the short-wavelength regime, causing an additional current and the associated electric field as

$$\boldsymbol{E} = -\frac{1}{en}\boldsymbol{J} \times \boldsymbol{B}. \tag{35}$$

The Hall term was expected to not contribute to the EMF in the case of isotropic forcing to turbulence in a uniform large-scale magnetic field on the basis of anti-symmetric properties of the velocity tensor in the wavevector domain (Gimblett and Allan 1976). However, direct numerical simulations of Hall-MHD show that the Hall current can strongly enhance or suppress generation of large-scale magnetic field energy, depending on the spatial scales compared to the system size (Mininni et al. 2003). Direct numerical simulations also show that the Hall current has the dual energy transfer character around the Hall scale (ion scale), providing both forward energy cascade to smaller scales and inverse energy cascade to larger scales (Mininni et al. 2007). In plasma physics sense, the Hall current can compete against non-fluid, non-Hall current such as the diamagnetic current (Narita et al. 2020).

### 2.4.3 Nonlinear effect

The nonlinearity of the EMF with respect to the large-scale magnetic field is proposed by considering third-order nonlinearity representing the (magnetic) ponderomotive force $\nabla(B^2)$ and the current helicity $\boldsymbol{J} \cdot \boldsymbol{B}$

$$\mathcal{E} = \delta_1 \nabla \times \boldsymbol{B} + \delta_2 (\nabla B^2) \times \boldsymbol{B} + \delta_3 (\nabla \times \boldsymbol{B} \cdot \boldsymbol{B})\boldsymbol{B}. \tag{36}$$

Equation (36) was proposed for a weak-field, axisymmetric turbulence case (Roberts and Stix 1971; Gimblett and Allan 1976) The transport coefficients $\delta_1$, $\delta_2$, and $\delta_3$ are expected to depend on the fluid properties. the spectrum of the turbulence, and the strength of the large-scale field.

### 2.5 Level of model simplification

Various EMF models described above can be characterized by different level of simplication. It is possible to distinguish those models from the simplified one to the elaborated ones in a systematic fashion by introducing the criterion as below. Under a proper decomposition between the mean (large-scale, resolved scale) and fluctuating (small-scale, unresolved scale) components, any closure formulation leads to a similar functional dependence on the mean fields from the fundamental equations. On the other hand, the expressions of the turbulent transport coefficients, such as $\alpha$, $\beta$, etc., can be fairly different depending on the closure scheme adopted. The simplest EMF expression (Eq. 5) with the transport coefficient expression (Eqs. 7 and 8) is obtained with the quasi-linear or the first-order smoothing approximation (FOSA) on the assumption of a low magnetic Reynolds number and neglecting the inhomogeneous mean velocity effect.

It should be noted here that the low magnetic Reynolds number is rather unlikely in space and astrophysical plasmas due to the large spatial scale and the collisionless or nearly collisionless property associated with the low density.

If we relax the assumption of low magnetic Reynolds number, and start considering the role of Lorentz force on the momentum equation, the current helicity contribution enters the $\alpha$ expression as $h_{\text{crt}} = \langle \delta \mathbf{B} \cdot (\nabla \times \delta \mathbf{B}) \rangle$ as in Eq. (28). The cross-helicity contribution enters the EMF expression as Eq. (30) with the relaxation of neglect of the inhomogeneous mean velocity effect.

## 3 Electromotive fields in space plasmas

### 3.1 Order of magnitude

In the observational sense, the evaluation of the EMF is straightforward when the magnetic field data and the flow velocity data available with the caveat that the construction of the large-scale fields (magnetic field and flow velocity) is not unique but there are different choices such as the local averaging, the smoothing, and the low-pass filtering. The EMF is expressed in units of electric field. The choice of mV km$^{-1}$ is convenient when expressing the magnetic field in nT and the flow velocity in km s$^{-1}$ in the branch of space plasma physics. The magnetic field has a fluctuation amplitude of about 1–10 nT in the solar wind, and the flow velocity about 1–10 km s$^{-1}$, indicating that the typical amplitude of EMF is of the order of 1–100 mV km$^{-1}$.

Empirically speaking, the quiet solar wind has an EMF of about 10 mV km$^{-1}$, and the active solar wind with interplanetary shocks, coronal mass ejections, corotating interaction regions, and magnetic clouds reaches an EMF amplitude of about 1000 mV $km^{-1}$ or higher. Figure 1 displays the histogram of the EMF amplitudes (the peak values) during the interplanetary shock crossings observed by Helios-1 and Helios-2 in the inner heliosphere (Hofer and Bourdin 2019). The distribution has a maximum between 100 and 1000 mV km$^{-1}$, and extends to an amplitude up to about 20,000 mV km$^{-1}$. Moreover, Hofer and Bourdin (2019) showed that the peak values of the EMF radially decays at larger distances from the Sun approximately by a power law as $r^{-3/2}$ (here, $r$ denotes the distance from the Sun) and suggested that the EMF is a useful indicator to detect interplanetary shocks.

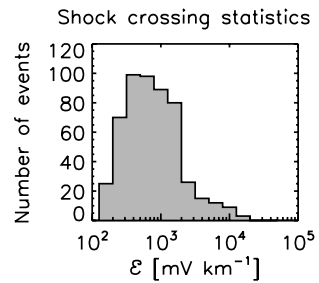

**Fig. 1** Histogram of EMF magnitude values during the interplanetary shock crossings observed by Helios-1 and Helios-2 in the inner heliosphere using the data in Hofer and Bourdin (2019)

### 3.2 Turbulent behavior

The EMF in space plasmas has a character of turbulent and random fluctuations. Marsch and Tu (1992, 1993) determined the frequency spectra of the EMF in the spacecraft frame when the Helios-1 spacecraft was located at a distance of about 0.53 astronomical units (between the Mercury orbit and the Venus orbit). The spectrum of the N component magnitude of the EMF (out-of-ecliptic, northward to the solar rotation axis, and perpendicular to the radial direction from the Sun) is shown in Fig. 2. The electromotive field oscillates the sign randomly in the turbulent solar wind, and the spectrum has a power-law curve with an index of about $-5/3$, like the power spectra for the magnetic field and the flow velocity in the turbulent solar wind.

The analyzed solar wind interval shows Alfvénic fluctuations with highly correlated variations between the magnetic field and the flow velocity. The flow speed was about 637 km s$^{-1}$. The frequency spectrum may be regarded as nearly streamwise wavenumber spectrum when Taylor's frozen-in flow hypothesis is used. The EMF vanishes in the purely Alfvénic fluctuations, since the fluctuating flow velocity is either positively or negatively correlated to the fluctuating magnetic field. The overall power-law spectral formation is indicative of some turbulent cascade mechanism operating in the EMF. On the other hand, the EMF profile is different during the times of shock or transient crossings, which is discussed in the subsection of transport coefficients.

### 3.3 Test for the mean-field dynamo model

Having the magnetic field data and the flow velocity data, one may test for the mean-field dynamo model against the solar wind observations in various ways. Historically, Marsch and Tu (1992) performed the pioneering work with the Helios data,

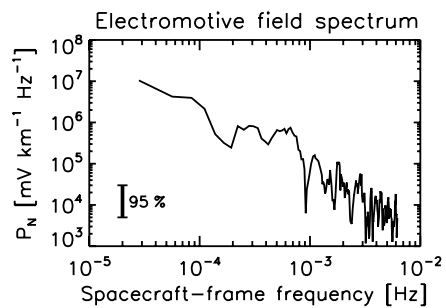

**Fig. 2** Spacecraft-frame frequency spectrum of the out-of-ecliptic component (the N component) of EMF using the Helios-1 data in 1980 at 0.53 astronomical units to the Sun. Spectral data are obtained by Marsch and Tu (1992). The magnitude of the EMF N component is plotted here

and examined the validity of the simplified alpha-term formula $\mathcal{E} \propto \alpha \boldsymbol{B}_0$. against the Helios-2 observation of fast solar wind around 0.29 AU on April 15–19, 1976 (hourly averaged data). Figure 3 displays a scatter plot of the EMF as a function of the three components (in the RTN coordinate system) of the mean magnetic field and the magnitude of the field. (the R component pointing radially away from the Sun, the N component pointing to the solar-ecliptic northward direction, and the T component completing the orthogonality and pointing to the azimuthal and westward direction). The alpha effect test shows no clear proportionality between the large-scale magnetic field and the EMF in the turbulent solar wind. Variation of the electromotive field is large, while the large-scale magnetic field does not vary much.

The proportionality test is revisited using a different analysis technique and a different Helios data set from that studied by Marsch and Tu (1992). The EMF is evaluated for a Helios-2 magnetic cloud crossing interval interval on April 17–20, 1978 at a distance of about 0.53 astronomical units to the Sun with a time resolution down to 40 s (Narita and Vörös 2018). The EMF is again compared both component-wise and magnitude-wise with the large-scale magnetic field (Fig. 4). Again, the scatter plot does not exhibit a clear proportionality between the EMF and the large-scale magnetic field. The R component is dominating the large-scale magnetic field (also representative of the magnitude panel), and the EMF varies over 4 orders of magnitude (in the range between 0.1 mV km$^{-1}$ and 1000 mV km$^{-1}$). The T and N components show nearly random distributions.

The lesson from the EMF studies by Marsch and Tu (1992) and Narita and Vörös (2018) is that there is no clear correlation with the large-scale magnetic field, indicating that the diffusion term (beta effect), the cross-helicity term, and higher-order terms may as well play an important role. Or the transport coefficient alpha cannot be regarded as nearly constant in the observational studies. Perhaps the proportionality can be better resolved in the Fourier domain, as the spatial derivative is replaced by the wavevectors, and further by the frequencies when employing Taylor's hypothesis.

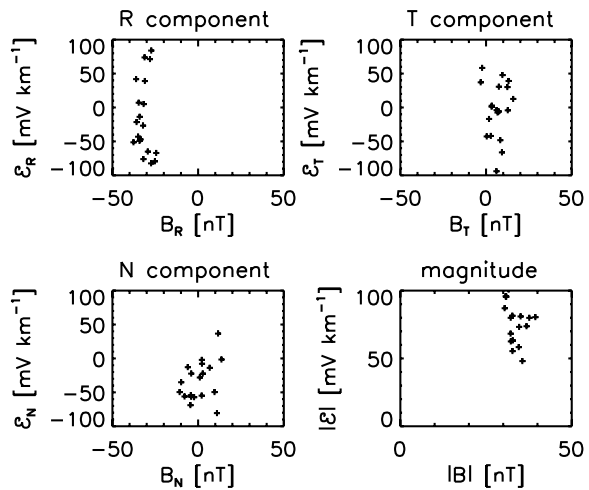

**Fig. 3** EMF evaluated in the component-wise and magnitude fashions as a function to the large-scale magnetic field. The Helios-2 magnetic field and ion data at a distance of 0.29 astronomical units from the Sun (April 15–19 1976) are used here (Marsch and Tu 1992). The measured time interval represents a high-speed solar wind with a speed of 733 km s$^{-1}$

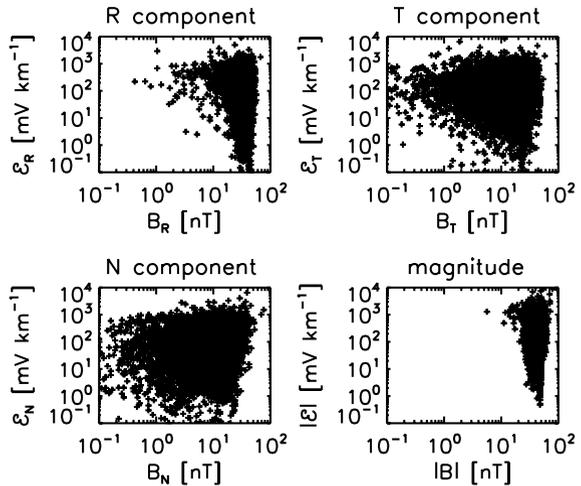

**Fig. 4** EMF evaluated in the same fashion as that in Fig. 3 for a Helios-2 time interval on April 17–20, 1978 at a heliocentric distance of about 0.53 astronomical units (Narita and Vörös 2018)

It should be noted here that there are different methods to decompose the observed field into the mean field and the fluctuation field. Local ensemble averaging (Marsch and Tu 1992; Narita and Vörös 2018) and moving Gaussian convolution (Bourdin et al. 2018) are used in the observational studies so far. Truncation (or band-passing) in the frequency domain into the low-frequency part as the mean field and the high-frequency part as the fluctuation part is also a competitive candidate of the field decomposition. The band-pass method strictly cuts the mixture between the low-frequency part and high-frequency part, but unlikely to Large Eddy Simulations (LES), the characteristic scale or frequency for the truncation is not known a priori and needs to be given by hand in the data analysis. The question remains open as to how much the EMF varies for these different field decomposition methods.

### 3.4 Transport coefficients

Another approach of testing for the mean-field dynamo model against the space plasma data is to evaluate the transport coefficients by the data inversion technique (Narita and Vörös 2018) or the reconstruction procedure (Bourdin et al. 2018).

In the data inversion approach, by limiting the analysis to the alpha and beta terms and treating the transport coefficients (alpha and beta) as independent from the large-scale field, one may derive the analytic estimators to evaluate the alpha and beta coefficients directly from the data (Narita and Vörös 2018; Narita 2021). The coefficients are then plotted as a function of the fluctuating flow velocity (Fig. 5 left panels) and the fluctuating magnetic field (Fig. 5 right panels). Interestingly, even though the level of data scattering is rather high, there is a weak indication or tendency that the transport coefficients depends on the fluctuating fields, such that the alpha and beta coefficients are larger at higher fluctuation amplitudes. This tendency appears both at the fluctuating flow velocity and the fluctuating magnetic field.

In the reconstruction procedure approach, a vortical flow motion is modeled and the analysis is extended to the transport coefficients of alpha, beta, and gamma for the cross-helicity effect (Bourdin et al. 2018). The EMF is evaluated as a time-series by directly computing from the fluctuating flow velocity data and the fluctuating magnetic field data (Fig. 6 data curve in black indicated by "observation") and compared with the reconstruction using the mean-field dynamo model (the cross-helicity dynamo model) using the large-scale fields and the vortex model (Fig. 6 data curve in gray indicated by "mean-field model"). The method is applied to the same shock crossing event as analyzed by Narita and Vörös (2018). The reconstruction of the EMF using the mean-field model gives a reasonable and qualitative fitting to the observation. The both methods highlight the EMF enhancing just at the shock crossing at about 1900 UT on April 18, 1978. The reconstruction method also gives the result that the gamma term (cross-helicity effect) reaches the same order of magnitude as the alpha term. While the picture of the mean-field dynamo might appear to fail in the proportionality tests in the turbulent solar wind (Figs. 3 and 4), Bourdin et al. (2018) successfully demonstrates that the mean-field model (even including the cross-helicity effect) can reasonably explain the observed EMF. Under what condition, the mean-field model works remains a question or motivation to further studies.

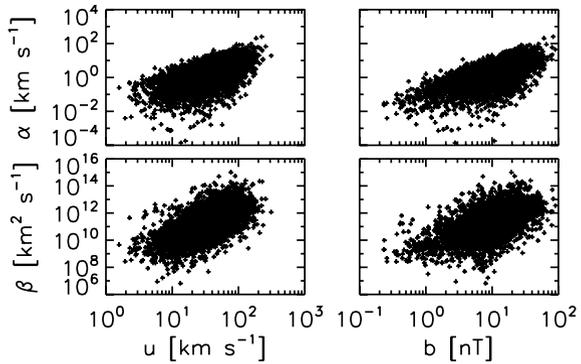

**Fig. 5** Transport coefficients (alpha and beta) evaluated as functions of the fluctuating flow velocity (left panels) and the fluctuating magnetic field (right panels); figure from Narita (2021)

### 3.5 Magnetospheric dipolarization fronts

When the magnetic field in the magnetotail of Earth's magnetosphere reconnects, this may lead to a subsequent reconfiguration of the global field structure. More precisely, the dipolar nature of the inner magnetosphere close to the Earth becomes then more dipolar. We call such events a dipolarization front and it can be observed with in-situ measurements from satellite observatories like ESA's *Cluster* or NASA's *MMS*. Typically, we find dipolarization fronts from the reconnection point inwards to the inner magnetosphere.

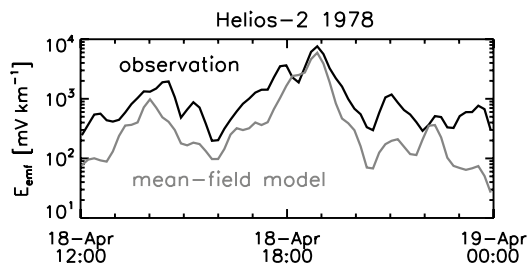

**Fig. 6** EMF determined directly from the fluctuation data (the flow velocity and the magnetic field, denoted by "observation") and that reconstructed using the cross-helicity dynamo model (denoted by "mean-field model"). Adapted from the study by Bourdin et al. (2018)

Figure 7 shows one typical example of a dipolarization front observed by the *Cluster* instruments in the magnetic field component $B_z$ that is mostly parallel to the magnetic moment of Earth, as well as the radial proton velocity $u_x$ of the magnetospheric bulk plasma flow. A positive $B_z$ means that the magnetic field is roughly aligned with the pre-existing dipolar field component, so that this front enhances the magnetic dipole moment, as expected. A positive $u_x$ means the bulk plasma flow, which the protons follow, is toward the planet. This is consistent with a plasma outflow stream from a reconnection site located further out in the magnetotail of Earth.

The EMF shown in the upper panel of Fig. 7 clearly features a significant peak, where the EMF rises from a background level of below 0.5 V/Mm by one order of magnitude to about 5 V/km. In other cases, we sometimes see electromotive-force peaks of about $10 - 20$ V/km. Once the main peak passes the spacecraft, we see that for a time of about 3 days, the level of the EMF remains significantly enhanced; see smoothed black solid line in the upper panel of Fig. 7.

The reason why we plot the $u_x$ and $B_z$ components here is that the mean-field EMF formulation consists partly of the cross product of the fluctuations of these quantities: $\delta u_x \times \delta B_z$. Still, we learn here that these are not the only important components, as the main peak consists with a peak in $B_z$, but the second peak that occurs about 12 hours later with over 3 V/km is caused neither by strong oscillations in $B_z$ nor in $u_x$. Future investigations of similar events using the mean-field formulation of the EMF will reveal more details on its generating mechanisms.

### 3.6 Interplanetary shock fronts

Coronal mass ejections (CMEs) are ejected form the Sun into the inner heliosphere and it is expected that these faster supersonic flows run into a slower but also supersonic background solar wind. Due to the so-called snow-plow effect, significant shock fronts develop and travel large distances in the solar system, even beyond Jupiter orbit. Observatories like ESA's *SolarOrbiter* or NASA's *Parker Solar Probe* are targeting to observe such interplanetary CMEs (iCMEs).

Recent in-situ observations of an iCME allow us to use both, the plasma bulk flow velocity and the magnetic field vector, to compute the EMF for the first time in high resolution. Previously, we used lower resolution data from Helios to evaluate statistical samples of iCMEs (Hofer and Bourdin 2019).

In Fig. 8, we analyze now *SolarOrbiter* measurements of 3 November 2021 around 1800 UT, while an iCME shock front passes over the spacecraft. Again, we see how the EMF makes a significant jump over one order of magnitude at the time of the shock-front arrival; see top panel in Fig. 8. To visualize the general trends, we average all resulting signals over 300 s. We see a sharp rise of the total solar-wind bulk velocity at the shock front; see middle panel of Fig. 8. The radial component $u_R$ is mostly similar to the absolute value of $u$ and is therefore omitted here.

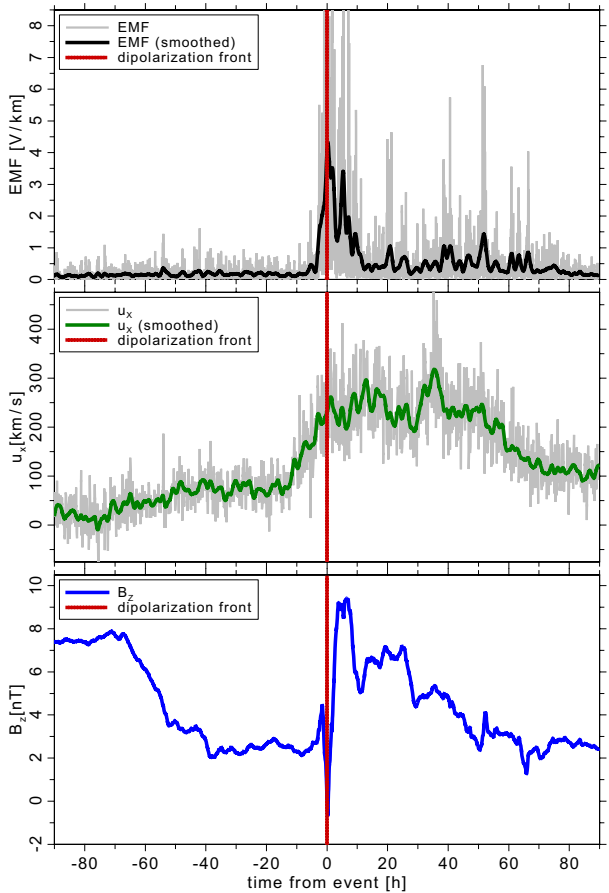

**Fig. 7** EMF (top panel) during a dipolarization front, observed onboard of *Cluster* in the inner magnetosphere of Earth. The planets's center is located in the positive x direction from this in-situ observation, where x is similar to the radial direction. The plasma bulk velocity (middle panel) is denoted as $u$ here and the high-resolution data are visible as gray background lines, which the green solid line shows a smoothing over 300 s. The magnetic field component $B_z$ (bottom panel) is roughly perpendicular to the ecliptic

The EMF remains strongly enhanced for about 12 h after the shock-front arrival. The return to the pre-event values can be as late as 2 days after the shock front. We think the EMF may play an important role in the deceleration and decay of iCMEs—or their lack of. The prediction of iCME arrival times at earth orbit might be improved with a better understanding of the impact of the enhanced EMF on the iCME drag forces (Amerstorfer et al. 2018).

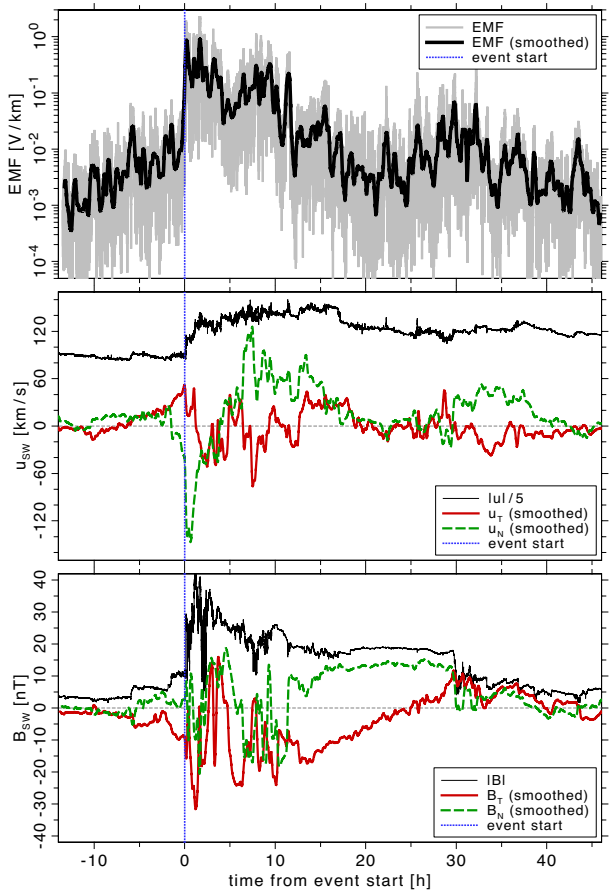

**Fig. 8** EMF during an interplanetary shock front in the solar wind (top panel), observed 3 November 2021 around 18:00 UT onboard of *SolarOrbiter* in the inner heliosphere. The tangential and normal components of the solar-wind velocity vector (middle panel), $u_T$ and $u_N$, are smoothed over 300 s, while the absolute value of $u$ is scaled down by a factor of 5 for better visibility. The absolute value of the magnetic field $B$ (bottom panel) is shown at the same scale as the tangential and radial components, $B_T$ and $B_N$

## 4 Outlook

### 4.1 Interplanetary space

The EMF has largely been overlooked in space and astrophysical plasma studies, while it is a key concept to describe the magnetic field generation (dynamo mechanism) and the turbulent fields in the plasma. The theory of EMF is still in the growing phase. Being the cross product between the flow velocity and the magnetic field, the EMF is one of the second-order magnetohydrodynamic (MHD) quantities according to the definition, but there is a possibility of renormalizing the field down to the first-order quantity (as a function of the magnetic field or the flow velocity) for the closure of the set of turbulent MHD equations. How the closure is realized is one of the urgent tasks for the space and astrophysical plasma communities.

The EMF is observable in near-Earth space (interplanetary space). The EMF is non-zero even in the quiet solar wind, and becomes enhanced when a transient event such as an interplanetary shock is passing by the spacecraft. The EMF magnitude varies from the order of mV km$^{-1}$ (quiet solar wind) to 1000 mV km$^{-1}$ (transient).

The EMF shows both random oscillations around the zero value and systematic behaviors when the field is enhanced at the shock crossing. For the random fluctuation parts, the power spectral analysis indicates that the EMF motion is reminiscent of plasma turbulence characterized by a power-law spectrum with an index of $-5/3$. For the systematic trends, not only the alpha and beta effects but also the cross-helicity effect (the gamma term) plays an important role to successfully reproduce the observed EMF profile.

### 4.2 Astrophysical plasmas

Different types of EMF generation may operate in astrophysical systems. Applications are highlighted at the interstellar medium with density and temperature gradients as well as the relativistic jets driven by the rotating black holes.

#### 4.2.1 Interstellar medium

Cross-field diffusion may lead to the non-turbulent EMF generation in interstellar space. The currents are carried by diffusing motion of electrons, and the excitation of the EMF (given as a curl of the EMF) is assessed as a vector product between the gradient of the ion temperature $T_i$ and that of the electron density $n_e$ (Lazarian 1992)

$$\nabla \times \boldsymbol{E} = \frac{4k_B}{\pi e n_e} \nabla T_i \times \nabla n_e. \tag{37}$$

Here, $k_B$ denotes the Boltzmann constant, and $e$ electric charge. Application of the cross-field diffusion scenario to the interstellar medium indicates a magnetic field of about $3 \times 10^{-17}$ G on a time scale of $10^9$ years. While a field strength of $10^{-17}$ may sound like a weak field, but it is sufficient to serve as a seed field in the galaxy for further dynamo mechanisms (Rees 1987; Lazarian 1992).

#### 4.2.2 Relativistic jets

EMFs are expected to play a crucial role in the relativistic jet formation (highly collimated outflow) by extracting the rotation energy of the black holes such as in the active galactic nuclei, the microquasars, and the gamma-ray bursts. Various energy conversion mechanisms have been proposed to explain the relativistic jets: the electromagnetic extraction of the rotational energy of the central black hole (Blandford and Znajek 1977), the electromagnetic extraction of the rotational energy of an accretion flow around the black hole (Lovelace 1976), and the thermal outflow from the accretion flow (Paczynski 1990). By regarding the electromagnetic field energy density dominates the black hole, the general-relativistic treatment of the Blandford-Znajek process yields the result that the EMF (including the permittivity) is excited in the radial direction to the central black hole (using the rotation in the azimuthal direction and the poloidal magnetic field) around of the rotating black hole (the ergosphere of Kerr black holes). The EMF is stronger (in terms of the energy density) than the magnetic field, and drives the current in the poloidal direction across the magnetic field, which in turn generates the magnetic field component in the azimuthal direction, serving as the collimation mechanism of the jets (Toma and Takahara 2014).

**Acknowledgements** This research was funded in whole or in part the Austrian Science Fund (FWF) [10.55776/P32958] and [10.55776/P37265]. For open access purpose, the author has applied a CC BY public copyright license to any author-accepted manuscript version arising from this submission.

**Funding** Open Access funding enabled and organized by Projekt DEAL.

**Data availability** Solar Orbiter and Cluster data are available at CDAWeb.

### Declarations

**Conflict of interest** The author declares no conflict of interest.

### References

T. Amerstorfer, C. Möstl, P. Hess, M. Temmer, M.L. Mays, M.A. Reiss, P. Lowrance, P.-A. Bourdin, Ensemble prediction of a halo coronal mass ejection using jeliospheric imagers. Space Weather **16**, 784–801 (2018). https://doi.org/10.1029/2017SW001786

R. Beck, L. Chamandy, E. Elson, E.G. Blackman, Synthesizing observations and theory to understand galactic magnetic fields: Progress and challenges. Galaxies **8**, 4 (2020). https://doi.org/10.3390/galaxies8010004

S.V. Berdyugina, Starspots: A key to the stellar dynamo. Living Rev. Solar Phys. **2**, 8 (2005). https://doi.org/10.12942/lrsp-2005-8

M.A. Berger, G.B. Field, The topological properties of magnetic helicity. J. Fluid Mech. **147**, 133–148 (1984). https://doi.org/10.1017/S0022112084002019

M.A. Berger, Introduction to magnetic helicity. Plasma Phys. Control. Fusion **41**, B167–B175 (1999). https://doi.org/10.1088/0741-3335/41/12B/312

R.D. Blandford, R.L. Znajek, Electromagnetic extraction of energy from Kerr black holes. Mon. Not. Roy. Astron. Soc. **179**, 433–456 (1977). https://doi.org/10.1093/mnras/179.3.433

P.-A. Bourdin, B. Hofer, Y. Narita, Inner structure of CME shock fronts revealed by the electromotive force and turbulent transport coefficients in Helios-2 observations. Astrophys. J. **855**, 111 (2018). https://doi.org/10.3847/1538-4357/aaae04

A. Brandenburg, Advances in mean-field dynamo theory and applications to astrophysical turbulence. J. Plasma Phys. **84**, 735840404 (2018). https://doi.org/10.1017/S0022377818000806

A.S. Brun, M.K. Browning, Magnetism, dynamo action and the solar-stellar connection. Living Rev. Solar Phys. **14**, 4 (2017). https://doi.org/10.1007/s41116-017-0007-8

P. Charbonneau, Dynamo models of the solar cycle. Living Rev. Solar Phys. **7**, 3 (2010). https://doi.org/10.12942/lrsp-2010-3

P. Charbonneau, Solar dynamo theory. Annu. Rev. Astron. Astrophys. **52**, 251–290 (2014). https://doi.org/10.1146/annurev-astro-081913-040012

W.M. Elsasser, Hydromagnetic dynamo theory. Rev. Mod. Phys. **28**, 135–163 (1956). https://doi.org/10.1103/RevModPhys.28.135

C.G. Gimblett, D.W. Allan, The electromotive force generated by driven plasma motions. J. Plasma Phys. **16**, 389–398 (1976). https://doi.org/10.1017/S0022377800020298

G.A. Glatzmaier, P.H. Roberts, Dynamo theory then and now. Int. J. Eng. Sci. **36**, 1325–1338, (1998). https://doi.org/10.1016/S0020-7225(98)00035-4

G.A. Glatzmaier, Geodynamo simulations - How realistic are they? Annu. Rev. Earth Planet. Sci. **30**, 237–257 (2002). https://doi.org/10.1146/annurev.earth.30.091201.140817

A.V. Gruzinov, P.H. Diamond, Self-consistent theory of mean-field electrodynamics. Phys. Rev. Lett. **72**, 1651 (1994). https://doi.org/10.1103/PhysRevLett.72.1651

H. Hamabata, T. Namikawa, Y. Hosoya, The mean electromotive force generated by random Alfvén waves in a collisionless non-uniform plasma. J. Plasma Phys. **28**, 309–315 (1982). https://doi.org/10.1017/S0022377800000295

H. Hamabata, T. Namikawa, Mean-field magnetohydrodynamics associated with random Alfvén waves in a plasma with weak magnetic diffusion. J. Plasma Physa. **39**, 139–149 (1988). https://doi.org/10.1017/S0022377800012903

J.F. Hawley, C.F. Gammie, S.A. Balbus, Local three-dimensional magnetohydrodynamic simulations of accretion disks. Astrophys. J. **440**, 742–763 (1995). https://doi.org/10.1086/175311

B. Hofer, P.-A. Bourdin, Application of the electromotive force as a shock front indicator in the inner heliosphere. Astrophys. J. **878**, 30 (2019). https://doi.org/10.3847/1538-4357/ab1e48

C.A. Jones, Planetary magnetic fields and fluid dynamics. Annu. Rev. Fluid Mech. **43**, 583–614 (2011). https://doi.org/10.1146/annurev-fluid-122109-160727

M. Kono, P.H. Roberts, Recent geodynamo simulations and observations of the geomagnetic field. Rev. Geophys. **40**, 4 (2002). https://doi.org/10.1029/2000RG000102

F. Krause, K.-H. Rädler, Mean-field Magnetohydrodynamics and Dynamo Theory, Pergamon. Oxford (1980). https://doi.org/10.1016/C2013-0-03269-0

P.P. Kronberg, Extragalactic magnetic fields. Rep. Prog. Phys. **57**, 325–382 (1994). https://doi.org/10.1088/0034-4885/57/4/001

R.M. Kulsrud, S.W. Anderson, The spectrum of random magnetic fields in the mean field dynamo theory of the galactic magnetic field. Astrophys. J. **396**, 606 (1992). https://doi.org/10.1086/171743

A. Lazarian, Diffusion-generated electromotive force and seed magnetic field problem. Astron. Astrophys. **264**, 326–330 (1992). https://articles.adsabs.harvard.edu/pdf/1992A%26A...264..326L

G. Lesur, G.I. Ogilvie, On self-sustained dynamo cycles in accretion discs. Astron. Astrophys. **488**, 451 (2008). https://doi.org/10.1051/0004-6361:200810152

R.V.E. Lovelace, Dynamo model of double radio sources. Nature **262**, 649–652 (1976). https://doi.org/10.1038/262649a0

E. Marsch, C.-Y. Tu, Electric field fluctuations and possible dynamo effects in the solar wind, Solar Wind Seven, Proceedings of the 3rd COSPAR Colloquium, Goslar, Germany, 16–20 September 1991, (eds) E. Marsch, R. Schwenn, Pergamonn Press, Oxford, pp. 505–510 (1992). https://doi.org/10.1016/B978-0-08-042049-3.50105-8

E. Marsch, C.-Y. Tu, MHD turbulence in the solar wind and interplanetary dynamo effects, The Cosmic Dynamo, Proceedings of the 157th Symposium of the International Astronomical Union held in Potsdam, Germany, 7–11 September 1992 (eds) F. Krause, K.-H. Rädler, G. Rüdiger, International Astronomical Union, Springer, Dordrecht, pp. 51–57 (1993). https://doi.org/10.1007/978-94-011-0772-3_9

P.D. Mininni, D. Gómez, S.M. Mahajan, Dynamo action in magnetohydrodynamics and Hall-magnetohydrodynamics. Astrophys. J. **587**, 472–481 (2003). https://doi.org/10.1086/368181

P.D. Mininni, A. Alexakis, A. Pouquet, Energy transfer in Hall-MHD turbulence: cascades, backscatter, and dynamo action. J. Plasma Phys. **73**, 377–401 (2007). https://doi.org/10.1017/S0022377806004624

H.K. Moffatt, The mean electromotive force generated by turbulence in the limit of perfect conductivity. J. Fluid Mech. **65**, 1–10 (1974). https://doi.org/10.1017/S0022112074001200

H.K. Moffatt, *Magnetic Field Generation in Electrically Conducting Fluids* (Cambridge University Press, Cambridge, 1978)

H.K. Moffatt, Helicity and singular structures in fluid dynamics. Proc. Nat. Acad. Sci. **111**, 3663–3670 (2014). https://doi.org/10.1073/pnas.1400277111

T. Namikawa, H. Hamabata, The mean electromotive force generated by random Alfvén waves in a collisionless plasma. J. Plasma Phys. **27**, 415–425 (1982). https://doi.org/10.1017/S0022377800010989

T. Namikawa, H. Hamabata, The mean electromotive force generated by random Alfvén waves in a collisionless plasma under a non-uniform mean magnetic field. J. Plasma Phys. **28**, 293–298 (1982). https://doi.org/10.1017/S0022377800000271

T. Namikawa, H. Hamabata, Y. Hosoya, The mean electromotive force generated by random hydromagnetic waves in a collisionless plasma. J. Plasma Phys. **28**, 299–307 (1982). https://doi.org/10.1017/S0022377800000283

T. Namikawa, H. Hamabata, The effect of microscale random Alfvén waves on the propagation of large-scale Alfvén waves. J. Plasma Phys. **29**, 243–253 (1983). https://doi.org/10.1017/S0022377800000738

T. Namikawa, H. Hamabata, The $\alpha$-effect generated by standing Alfvén waves. J. Plasma Phys. **40**, 353–358 (1988). https://doi.org/10.1017/S0022377800013325

Y. Narita, Z. Vörös, Evaluation of electromotive force in interplanetary space. Ann. Geophys. **36**, 101–106 (2018). https://doi.org/10.5194/angeo-36-101-2018

Y. Narita, O. W. Roberts, Z. Vörös, M. Hoshino, Transport ratios of the kinetic Alfvén mode in space plasmas. Front. Phys. **8**, 166 (2000). https://doi.org/10.3389/fphy.2020.00166

Y. Narita, Electromotive force in the solar wind. Ann. Geophys. **39**, 759–768 (2021). https://doi.org/10.5194/angeo-39-759-2021

B. Paczynski, Super-Eddington winds from neutron stars. Astrophys. J. **363**, 218 (1990). https://doi.org/10.1086/169332

E. N. Parker, Cosmical Magnetic Fields, The International Series of Monographs on Physics (1979)

A. Pouquet, U. Frisch, J. Léorat, Strong MHD helical turbulence and the nonlinear dynamo effect. J. Fluid Mech. **77**, 321–354 (1976). https://doi.org/10.1017/S0022112076002140

K.-H. Rädler, N. Kleeorin, I. Rogachevskii, The mean electromotive force for MHD turbulence: The case of a weak mean magnetic field and slow rotation. Geophys. Astron. Fluid Dyn. **97**, 249–274 (2003). https://doi.org/10.1080/0309192031000151212

K.-H. Rädler, R. Stepanov, Mean electromotive force due to turbulence of a conducting fluid in the presence of mean flow. Phys. Rev. E **73**, 056311 (2006). https://doi.org/10.1103/PhysRevE.73.056311

K.-H. Rädler, M. Rheinhardt, Mean-field electrodynamics: critical analysis of various analytical approaches to the mean electromotive force. Geophys. Astrophys. Fluid Dyn. **101**, 117–154 (2007). https://doi.org/10.1080/03091920601111068

M. J. Rees, The origin and cosmogonic implications of seed magnetic fields. Quart. J. Roy. Astron. Soc. **28**, 197–206 (1971). https://articles.adsabs.harvard.edu/pdf/1987QJRAS..28..197R

P. H. Roberts, M. Stix, The turbulent dynamo: A translation of a series of papers by F. Krause, K.-H Räadler, and M. Steenbeck (No. NCAR/TN-60+IA). University Corporation for Atmospheric Research (1971). https://doi.org/10.5065/D6DJ5CK7

P.H. Roberts, A.M. Soward, Dynamo theory. Ann. Rev. Fluid Mech. **24**, 459–512 (1992). https://doi.org/10.1146/annurev.fl.24.010192.002331

P. Roberts, G.A. Glatzmaier, Geodynamo theory and simulations. Rev. Mod. Phys. **72**, 1081–1123 (2000). https://doi.org/10.1103/RevModPhys.72.1081

I. Rogachevskii, N. Kleeorin, Electromotive force and large-scale magnetic dynamo in a turbulent flow with a mean shear. Phys. Rev. E **68**, 036301 (2003). https://doi.org/10.1103/PhysRevE.68.036301

G. Rüdiger, M. Küker, The influence of helical background fields on current helicity and electromotive force of magnetoconvection. Astron. Astrophys. **592**, A73 (2016). https://doi.org/10.1051/0004-6361/201527145

G.R. Sarson, C.A. Jones, K. Zhang, G. Schubert, Magnetoconvection dynamos and the magnetic fields of Io and Ganumede. Science **276**, 1106–1108 (1997). https://doi.org/10.1126/science.276.5315.1106

G. Schubert, K. Zhang, M.G. Kivelson, J.D. Anderson, The magnetic field and internal structure of Ganymede. Nature **384**, 544–545 (1996). https://doi.org/10.1038/384544a0

N. Seehafer, Current helicity and the turbulence electromotive force. Europhys. Lett. **27**, 353–357 (1994). https://doi.org/10.1209/0295-5075/27/5/004

J. Squire, A. Bhattacharjee, Electromotive force due to magnetohydrodynamic fluctuations in sheared rotating turbulence. Phys. Rev. E **92**, 053101 (2015). https://doi.org/10.1103/PhysRevE.92.053101

M. Steenbeck, F. Krause, K.-H. Rädler, Berechnung der mittleren Lorentz-Feldstärke für ein elektrisch leitendes Medium in turbulenter, durch Coriolis-Kräfte beeinflußter Bewegung. Z. Naturforsch. **21a**, 369–376 (1966). https://doi.org/10.1515/zna-1966-0401

K. Toma, F. Takahara, Electromotive force in the Blandford-Znajek process. Mon. Not. R. Astron. Soc. **442**, 2855–2866 (2014). https://doi.org/10.1093/mnras/stu1053

V. Urpin, Mean electromotive force and dynamo action in a turbulent flow. Astron. Astrophys. **347**, L47–L50 (1999). https://adsabs.harvard.edu/full/1999A%26A...347L..47U

V. Urpin, Mean electromotive force in turbulent shear flow. Phys. Rev. E **65**, 026301 (2002). https://doi.org/10.1103/PhysRevE.65.026301

S.I. Vainshtein, A. A. Ruzmaikin, Generation of the large-scale galactic magnetic field. Astron. Z. **48**, 902–909 (1971). (English translation in Sov. Astron. AJ, 15, 714–719, 1972) https://adsabs.harvard.edu/full/1972SvA....15..714V

L.M. Widrow, Origin of galactic and extragalactic magnetic fields. Rev. Mod. Phys. **74**, 775–823 (2002). https://doi.org/10.1103/RevModPhys.74.775

N. Yokoi, G. Balarac, Cross-helicity effects and turbulent transport in magnetohydrodyamic flow. J. Phys: Conf. Ser. **318**, 072039 (2011). https://doi.org/10.1088/1742-6596/318/7/072039

N. Yokoi, Cross helicity and related dynamo. Geophys. Astrophys. Fluid Dyn. **107**, 114–184 (2013). https://doi.org/10.1080/03091929.2012.754022

N. Yokoi, Electromotive force in strongly compressible magnetohydrodynamic turbulence. J. Plasma Phys. **84**, 735840501 (2018). https://doi.org/10.1017/S0022377818000727

N. Yokoi, S. M. Tobias, Magnetoclinicity instability. In: R. Örlü, A. Talamelli, J. Peinke, M. Oberlack (eds) Progress in Turbulence IX, Proceedings of the iTi Conference in Turbulence 2021. Springer Proceedings in Physics, vol. 267, pp. 273–279 (2021). https://doi.org/10.1007/978-3-030-80716-0_37

A. Yoshizawa, Self consistent turbulent dynamo modeling of reversed field pinches and planetary magnetic fields. Phys. Fluids B **2**, 1589–1600 (1990). https://doi.org/10.1063/1.859484